\documentclass[onecolumn]{aastex63}
\usepackage{amsmath}
\usepackage{amssymb}

\submitjournal{ApJ}
\shorttitle{Nonlinear Wave Mixing}
\shortauthors{Jermyn}
\begin{document}

\title{Nonlinear Mixing driven by Internal Gravity Waves}

\correspondingauthor{Adam S. Jermyn}
\email{adamjermyn@gmail.com}

\author[0000-0001-5048-9973]{Adam S. Jermyn}
\affiliation{Center for Computational Astrophysics, Flatiron Institute, New York, NY 10010, USA}

\begin{abstract}
Hydrodynamic waves propagate through stellar interiors, transporting energy and angular momentum. They can also advect fluid elements to produce mixing, but this effect has not been quantified from first principles. We derive the leading order non-linear wave mixing due to internal gravity waves in a thermally and compositionally-stratified fluid. We find that this scales as the fourth power of wave velocity, that it is suppressed by compositional stratification, and that it depends on the thermal and compositional diffusivities.
\end{abstract}


\section{Introduction}

Waves are solutions to the linearized equations of motion of a fluid.
These are exact solutions to the full equations of motion in the limit of vanishing amplitude, but at any finite amplitude there are non-linear corrections.
One such correction is the Stokes Drift~\citep{andrews_mcintyre_1978}, which is the difference between the Eulerian displacement
\begin{align}
	\boldsymbol{\xi}_{\rm Euler}(\boldsymbol{r},t) \equiv \int_0^{t} \boldsymbol{u}(\boldsymbol{r}) dt
\end{align}
and the Lagrangian one
\begin{align}
	\boldsymbol{\xi}_{\rm Lagrange} \equiv \int_0^{t} \boldsymbol{u}(\boldsymbol{\xi}_{\rm Lagrange}(\boldsymbol{r},t)) dt
\end{align}
after some amount of time $t$.
That is,
\begin{align}
	\boldsymbol{\xi}_{\rm Stokes} = \boldsymbol{\xi}_{\rm Lagrange} - \boldsymbol{\xi}_{\rm Euler}
\end{align}
Here $\boldsymbol{u}$ is the velocity, $\boldsymbol{r}$ is the spatial coordinate, $t$ is time, and $\boldsymbol{\xi}$ is the displacement.

Here our aim is to derive the diffusivity associated with the Stokes drift for a random field of internal gravity waves (IGW).
Our approach is intentionally didactic, and we reproduce a number of known intermediate results for clarity.

We begin in Section~\ref{sec:diff} with a review of properties of diffusion, concluding with a well-known expression for the diffusion coefficient in terms of the zero-frequency autocorrelation of the velocity field.
In Section~\ref{sec:igw} we derive the equations of motion for internal gravity waves in a thermally- and compositionally-stratified medium, retaining both thermal and compositional diffusion.
We then use this to derive the non-linear forcing due to IGW, and use that to compute the nonlinear wave diffusivity.
We conclude with a comparison to other prescriptions for wave mixing in Section~\ref{sec:conclusions}.

\section{Diffusion}\label{sec:diff}

We now review some basic facts about diffusion.

\subsection{Diffusivity}

The diffusivity $D$ is defined in one dimension as
\begin{align}
D \equiv \lim_{T\rightarrow \infty} \frac{\langle (r(T)-r(0))^2\rangle}{2 T},
\end{align}
where $r(T)$ is the coordinate of a particle at time $T$ undergoing stochastic motion and $\langle … \rangle$ represents an expectation value over that motion.
We can relate this form to the velocity $u(t)$ of the particle via
\begin{align}
r(T) - r(0) = \int_0^T u(t) dt,
\end{align}
which gives
\begin{align}
D = \lim_{T\rightarrow \infty} \frac{1}{2T}\int_0^T \int_0^T \langle u(t) u(t')\rangle dt dt'.
\label{eq:Duu}
\end{align}

\subsection{Stationary Process}

In a stationary process correlations are time-translation invariant, so
\begin{align}
\langle u(t) u(t')\rangle = \langle u(t-t') u(0)\rangle.
\end{align}
This is a good approximation in most astrophysical contexts, where the forcing mechanism (e.g. convection) and the propagating medium change on time-scales which are very long compared with the wave frequency.

We proceed assuming that $u$ is described by a stationary process.
Equation~\eqref{eq:Duu} then simplifies to
\begin{align}
D = \lim_{T\rightarrow \infty} \frac{1}{2T}\int_0^T \int_{t-T}^{t} \langle u(\tau) u(0)\rangle d\tau dt,
\end{align}
where $\tau \equiv t-t'$.
Exchanging the order of integration we find
\begin{align}
D = \lim_{T\rightarrow \infty} \frac{1}{2T}\left(\int_{-T}^{0} \int_{0}^{T+\tau} + \int_{0}^{T} \int_{\tau}^{T}\right)\langle u(\tau) u(0)\rangle dt d\tau.
\end{align}
Once more because $u$ is a stationary process we have
\begin{align}
\langle u(\tau)u(0)\rangle = \langle u(0)u(-\tau)\rangle = \langle u(-\tau)u(0)\rangle,
\end{align}
so we can flip the sign of $\tau$ in the first pair of integrals and obtain
\begin{align}
D = \lim_{T\rightarrow \infty} \frac{1}{2T}\left(\int_{0}^{T} \int_{0}^{T-\tau} + \int_{0}^{T} \int_{\tau}^{T}\right)\langle u(\tau) u(0)\rangle dt d\tau
\end{align}
Performing the inner integrals over $t$ we find
\begin{align}
D = \lim_{T\rightarrow \infty} \int_{0}^{T} \frac{T-\tau}{T}\langle u(\tau) u(0)\rangle d\tau.
\end{align}
Taking the limit we recover the relation by~\citet{Kubo}:
\begin{align}
D = \int_{0}^{\infty} \langle u(\tau) u(0)\rangle d\tau,
\label{eq:Kubo}
\end{align}
which may also be written for a stationary process as
\begin{align}
D = \frac{1}{2} \int_{-\infty}^{\infty} \langle u(\tau) u(0)\rangle d\tau.
\label{eq:Kubo2}
\end{align}

\subsection{Relation to the Power Spectrum}

We use the Fourier transorm convention
\begin{align}
u(\omega) &= \int_{-\infty}^{\infty} e^{-i\omega t} u(t) \frac{dt}{\sqrt{2\pi}}\\
u(t) &= \int_{-\infty}^{\infty} e^{i\omega t} u(\omega) \frac{d\omega}{\sqrt{2\pi}}.
\end{align}
With this, we write the frequency autocorrelation as
\begin{align}
\langle u(\omega) u(\omega')\rangle = \int_{-\infty}^{\infty} \frac{dt}{\sqrt{2\pi}}\int_{-\infty}^{\infty} \frac{dt'}{\sqrt{2\pi}} e^{-i(\omega t + \omega' t')}\langle u(t) u(t')\rangle.
\end{align}
Because this is a stationary process we can subtract an offset from the times in the correlation function so long as the difference between them is preserved.
We do this with a change of variables to $\tau = t-t'$ and $q=(t+t')/2$, giving
\begin{align}
\langle u(\omega) u(\omega')\rangle &= \int_{-\infty}^{\infty} \frac{d\tau}{\sqrt{2\pi}}\int_{-\infty}^{\infty} \frac{dq}{\sqrt{2\pi}} e^{-i(\omega (\tau/2+q) - \omega' (\tau/2-q))}\langle u(\tau) u(0)\rangle\\
&= \int_{-\infty}^{\infty} \frac{d\tau}{\sqrt{2\pi}}\int_{-\infty}^{\infty} \frac{dq}{\sqrt{2\pi}} e^{-i\tau(\omega-\omega')/2-iq(\omega+\omega')}\langle u(\tau) u(0)\rangle\\
&= \delta(\omega+\omega')\int_{-\infty}^{\infty} d\tau e^{-i\tau(\omega-\omega')/2}\langle u(\tau) u(0)\rangle\\
&= \delta(\omega+\omega')\int_{-\infty}^{\infty} d\tau e^{-i\tau \omega}\langle u(\tau) u(0)\rangle\label{eq:finalUU},
\end{align}
where we obtained the third line by performing the integral over $q$.
Because the frequency autocorrelation vanishes except when $\omega=-\omega'$, we can define the power spectrum
\begin{align}
	S(\omega) \equiv \int_{-\infty}^{\infty} d\omega' \langle u(\omega) u(\omega')\rangle,
	\label{eq:S}
\end{align}
which is the energy per unit frequency in the velocity field.
Using equation~\eqref{eq:finalUU} we see that
\begin{align}
	S(\omega) = \int_{-\infty}^{\infty} d\tau e^{-i\tau \omega}\langle u(\tau) u(0)\rangle
\end{align}
and so
\begin{align}
	D = \frac{1}{2}S(0).
	\label{eq:DS0}
\end{align}
Hence, the diffusivity is related to the power spectrum at zero frequency.
Physically, this is because diffusion is a statement about long-time behaviour.

\subsection{Spatial Variation}

The diffusion coefficient is defined in terms of the motion of a single particle in the infinite-time limit, and so it is not trivial to define diffusion coefficients which vary in space.
It can be done, however, by defining the local diffusivity to be given by the diffusion coefficient one would obtain if the local conditions held globally.
We compute this by averaging the diffusion coefficient a volume $V$ which is small compared with the large-scale structure of the star but large compared with the characteristic length-scale of the velocity field (e.g. its scale of variation).
Thus we generalize equation~\eqref{eq:Kubo2} to find
\begin{align}
	D = \frac{1}{2} \int_{-\infty}^{\infty} d\tau \int \frac{d^3\boldsymbol{r}}{V} \langle u(\tau,\boldsymbol{r}) u(0,\boldsymbol{r})\rangle.
	\label{eq:Kubo3}
\end{align}
We now generalize our earlier Fourier transorm convention
\begin{align}
u(\omega,\boldsymbol{k}) &= \int_{-\infty}^{\infty} \frac{dt}{\sqrt{2\pi}}\int \frac{d^3\boldsymbol{r}}{V} e^{-i\omega t-i\boldsymbol{k}\cdot\boldsymbol{r}} u(t,\boldsymbol{r})\label{eq:uWK}  \\
u(t,\boldsymbol{r}) &= \int_{-\infty}^{\infty} \frac{d\omega}{\sqrt{2\pi}} \sum_{\boldsymbol{k}} e^{i\omega t+i\boldsymbol{k}\cdot\boldsymbol{r}} u(\omega,\boldsymbol{k})\label{eq:uTR},
\end{align}
as well as the corresponding mixed conventions for e.g. $u(\omega,\boldsymbol{r})$.

Defining
\begin{align}
	S(\omega,\boldsymbol{k}) \equiv \int_{-\infty}^{\infty} d\omega' \langle u(\omega,\boldsymbol{k}) u(\omega',-\boldsymbol{k})\rangle,
	\label{eq:S2}
\end{align}
we find
\begin{align}
	S(\omega,\boldsymbol{k}) &= \int_{-\infty}^{\infty} d\omega' \int \frac{d^3\boldsymbol{r}}{V}\int \frac{d^3\boldsymbol{r}'}{V} e^{i\boldsymbol{k}\cdot\boldsymbol{r} -i\boldsymbol{k}\cdot\boldsymbol{r}'}\langle u(\omega,\boldsymbol{r}) u(\omega',\boldsymbol{r'})\rangle
\end{align}
Using the results of the previous section we write this as
\begin{align}
	S(\omega,\boldsymbol{k}) &= \int_{-\infty}^{\infty} d\omega'\int_{-\infty}^{\infty} d\tau \int \frac{d^3\boldsymbol{r}}{V}\int \frac{d^3\boldsymbol{r}'}{V} e^{i\boldsymbol{k}\cdot\boldsymbol{r} -i\boldsymbol{k}\cdot\boldsymbol{r}' - i\omega \tau}\delta(\omega+\omega')\langle u(\tau,\boldsymbol{r}) u(0,\boldsymbol{r'})\rangle\\
	&= \int_{-\infty}^{\infty} d\tau \int \frac{d^3\boldsymbol{r}}{V}\int \frac{d^3\boldsymbol{r}'}{V} e^{i\boldsymbol{k}\cdot\boldsymbol{r} -i\boldsymbol{k}\cdot\boldsymbol{r}' + i\omega \tau}\langle u(\tau,\boldsymbol{r}) u(0,\boldsymbol{r'})\rangle
\end{align}
Summing over $\boldsymbol{k}$ produces $\delta(\boldsymbol{r}-\boldsymbol{r}')$, so
\begin{align}
	D = \frac{1}{2}\sum_{\boldsymbol{k}} S(0,\boldsymbol{k}).
	\label{eq:Kubo5}
\end{align}
That is, the diffusivity receives a contribution from the power at zero frequency for all wave-vectors.

\section{Internal Gravity Waves}\label{sec:igw}

\subsection{Leading Order}

Here we determine the leading order of the diffusivity in the wave velocity field $\boldsymbol{u}_w$.
This must be at least second order (i.e. $D \propto u_w^2$), as the diffusivity is sensitive to the power in the velocity field and hence goes like $u^2$.
However, the damping length of internal gravity waves approaches zero for internal gravity waves as $\omega \rightarrow 0$. So we should expect the power to vanish at $\omega=0$ anywhere away from the wave excitation region, and hence the contribution to the diffusivity vanishes as well.
This means that the dominant contribution to the diffusivity must arise at higher orders in $u_w$.
If we assume that the waves velocities are Gaussian random variables, then expectation values of the form $\langle u_w u_w u_w\rangle$ vanish, so the diffusivity must be at least fourth order in the wave velocity (i.e. $D \propto u_w^4$), and indeed some fourth-order terms do not straightforwardly vanish (specifically, terms in which each frequency occurs at least twice).

Fourth-order terms must arise via non-linear interactions between waves.
There are several terms in the Navier-Stokes equation that can provide such interactions, but the simplest is the Stokes acceleration
\begin{align}
\boldsymbol{a}_s = \boldsymbol{u}_w \cdot \nabla \boldsymbol{u}_w,
\label{eq:stokes}
\end{align}
Because $\boldsymbol{a}_s$ is quadratic in $\boldsymbol{u}_w$ it has power at zero frequency, even though $\boldsymbol{u}_w$ does not.
A quick way to see this is to note that $\sin(\omega t)$ has no power at zero-frequency (i.e. the Fourier transform has support only at $\omega'=\{-\omega,\omega\}$), but $\sin^2(\omega t)$, which has a non-zero time average, has support at $\omega'=\{-2\omega,0,2\omega\}$. The net result is that a substantial (order unity) fraction of the power in $a_{s,r}$ is at zero frequency.
While there are other non-linearities arising due to wave motion (e.g. coupling between the density and velocity fields), we believe that this term is representative of the largest of those and proceed neglecting all others.

We thus conclude that the diffusivity is most likely to arise in terms of the form $\langle \boldsymbol{u}_s \boldsymbol{u}_s\rangle$, where $\boldsymbol{u}_s$ is the velocity field that arises from a zero-frequency non-linear forcing term as derived below (see equation~\eqref{eq:stokes}).

That is, the waves interact with each other to produce a non-linear acceleration term that appears in the Navier-Stokes equations. This new term has a zero-frequency component, which drives further motion via the \emph{linear} equations of motion.
This new motion has a zero frequency component, and that is what enters into equation~\eqref{eq:DS0} to produce diffusion.

\subsection{A Subtlety with Wavevectors}

Because IGW are incompressible, the wavevector $\boldsymbol{k}$ obeys $\boldsymbol{k}\cdot\boldsymbol{u}_w = 0$, so equation~\eqref{eq:stokes} is non-zero only when there are multiple waves of different wave-vectors present.
This, however, is straightforward to arrange.
Consider for instance waves with wave-vectors $\boldsymbol{k}_r + \boldsymbol{k}_\perp$ and $\boldsymbol{k}_r - \boldsymbol{k}_\perp$.
These produces Stokes acceleration with terms of the form and magnitude
\begin{align}
\boldsymbol{a}_s &= (\boldsymbol{k}_r - \boldsymbol{k}_\perp)\cdot\boldsymbol{u}_w(\boldsymbol{k}_r + \boldsymbol{k}_\perp)\boldsymbol{u}_w(\boldsymbol{k}_r - \boldsymbol{k}_\perp).
\end{align}
Noting that $(\boldsymbol{k}_r+\boldsymbol{k}_\perp)\cdot \boldsymbol{u}_w(\boldsymbol{k}_r+ \boldsymbol{k}_\perp)=0$, we can rewrite this as
\begin{align}
\boldsymbol{a}_s &= \left[(\boldsymbol{k}_r - \boldsymbol{k}_\perp)-(\boldsymbol{k}_r + \boldsymbol{k}_\perp)\right]\cdot\boldsymbol{u}_w(\boldsymbol{k}_r + \boldsymbol{k}_\perp)\boldsymbol{u}_w(\boldsymbol{k}_r - \boldsymbol{k}_\perp)\\
&= -2\boldsymbol{k}_\perp\cdot\boldsymbol{u}_w(\boldsymbol{k}_r + \boldsymbol{k}_\perp)\boldsymbol{u}_w(\boldsymbol{k}_r - \boldsymbol{k}_\perp)\\
&= -2 k_\perp u_{w,\perp}(\boldsymbol{k}_r + \boldsymbol{k}_\perp) \boldsymbol{u}_w(\boldsymbol{k}_r - \boldsymbol{k}_\perp)\label{eq:stokes}
\end{align}
which is non-zero.

\subsection{Outline of Calculation}

We now sketch the calculation before performing it in more detail.

We first derive the linearized equations of motion for IGW in the Boussinesq plane-parallel limit.
We then apply the non-linear acceleration in equation~\eqref{eq:stokes} to those equations, and derive the linear response of the velocity field to the Stokes forcing.
The result is a radial velocity $u_r(\omega,\boldsymbol{k})$, where $\omega$ and $\boldsymbol{k}$ are the frequency and wavevector of $\boldsymbol{a}_s$.
This linear response is of the form
\begin{align}
	u_r(\omega,\boldsymbol{k}) = \mathcal{L}_i(\omega,\boldsymbol{k}) a_{s,i}(\omega,\boldsymbol{k}),
\end{align}
where $\mathcal{L}$ is a linear operator that depends on frequency and wavevector $\boldsymbol{k}$ and summation is implied over repeated indices.

Next, we relate the Stokes acceleration to the wave velocity field via equation~\eqref{eq:stokes}, which we write as
\begin{align}
	\boldsymbol{a}_s(\omega,\boldsymbol{k}) = \int_{-\infty}^{\infty} \frac{dt}{\sqrt{2\pi}}\int \frac{d^3\boldsymbol{r}}{V} e^{-i\omega t-i\boldsymbol{k}\cdot\boldsymbol{r}} \boldsymbol{u}_w(t,\boldsymbol{r}) \cdot \nabla \boldsymbol{u}_w(t,\boldsymbol{r})
\end{align}
Inserting equation~\eqref{eq:uTR} twice we find
\begin{align}
\boldsymbol{a}_s(\omega,\boldsymbol{k}) &= \int_{-\infty}^{\infty} \frac{dt}{\sqrt{2\pi}}\int \frac{d^3\boldsymbol{r}}{V} e^{-i\omega t-i\boldsymbol{k}\cdot\boldsymbol{r}} \int_{-\infty}^{\infty} \frac{d\omega'}{\sqrt{2\pi}} \sum_{\boldsymbol{k}'} e^{i\omega' t+i\boldsymbol{k}'\cdot\boldsymbol{r}} \boldsymbol{u}_w(\omega',\boldsymbol{k}')
 \cdot \nabla \int_{-\infty}^{\infty} \frac{d\omega''}{\sqrt{2\pi}} \sum_{\boldsymbol{k}''} e^{i\omega'' t+i\boldsymbol{k}''\cdot\boldsymbol{r}} \boldsymbol{u}_w(\omega'',\boldsymbol{k}'')\\
 &= \int_{-\infty}^{\infty} \frac{d\omega'}{\sqrt{2\pi}} \sum_{\boldsymbol{k}'} \int_{-\infty}^{\infty} \frac{d\omega''}{\sqrt{2\pi}} \sum_{\boldsymbol{k}''}\int_{-\infty}^{\infty} \frac{dt}{\sqrt{2\pi}}\int \frac{d^3\boldsymbol{r}}{V} e^{it(-\omega+\omega'+\omega'')+i\boldsymbol{r}\cdot(-\boldsymbol{k}+\boldsymbol{k}'+\boldsymbol{k}'')}   \boldsymbol{u}_w(\omega',\boldsymbol{k}')
 \cdot \boldsymbol{k}''\boldsymbol{u}_w(\omega'',\boldsymbol{k}'')\\
 &= \int_{-\infty}^{\infty} \frac{d\omega'}{\sqrt{2\pi}} \sum_{\boldsymbol{k}'}\boldsymbol{u}_w(\omega',\boldsymbol{k}')
 \cdot (\boldsymbol{k}-\boldsymbol{k}')\boldsymbol{u}_w(\omega-\omega',\boldsymbol{k}-\boldsymbol{k}').
\end{align}

The diffusivity is then given by
\begin{align}
	D &= \frac{1}{2}\sum_{\boldsymbol{k}} S(0,\boldsymbol{k})\\
	&= \frac{1}{2}\sum_{\boldsymbol{k}} \int_{-\infty}^{\infty} d\omega \langle u_r(0,\boldsymbol{k}) u_r(\omega,-\boldsymbol{k})\rangle\\
	&= \frac{1}{2} \int_{-\infty}^{\infty} d\omega \int_{-\infty}^{\infty} d\omega_1\int_{-\infty}^{\infty} d\omega_2 \sum_{\boldsymbol{k},\boldsymbol{k}_1,\boldsymbol{k}_2} \mathcal{L}_{a}(0,\boldsymbol{k})  \mathcal{L}_{c}(\omega,-\boldsymbol{k}) (\boldsymbol{k}-\boldsymbol{k}_1)_a (-\boldsymbol{k}-\boldsymbol{k}_2)_c \\
	&\times\langle(u_{w,a}(\omega_1,\boldsymbol{k}_1) u_{w,b}(-\omega_1,\boldsymbol{k}-\boldsymbol{k}_1) u_{w,c}(\omega_2,\boldsymbol{k}_2)u_{w,d}(\omega-\omega_2,-\boldsymbol{k}-\boldsymbol{k}_2)\rangle\label{eq:Dmess}.
\end{align}
The result is a four-point autocorrelation function of the wave field.

For Gaussian random variables $x_1...x_4$, Wick's theorem allows us to write
\begin{align}
	\langle x_1 x_2 x_3 x_4\rangle = \langle x_1 x_2\rangle \langle x_3 x_4\rangle + \langle x_1 x_3\rangle \langle x_2 x_4\rangle + \langle x_1 x_4\rangle \langle x_2 x_3\rangle.
\end{align}
We now approximate the correlations in $\boldsymbol{u}_w$ as Gaussian and use the above result to write
\begin{align}
&\langle(u_{w,a}(\omega_1,\boldsymbol{k}_1) u_{w,b}(-\omega_1,\boldsymbol{k}-\boldsymbol{k}_1) u_{w,c}(\omega_2,\boldsymbol{k}_2)u_{w,d}(\omega-\omega_2,-\boldsymbol{k}-\boldsymbol{k}_2)\rangle =\\
&\langle(u_{w,a}(\omega_1,\boldsymbol{k}_1) u_{w,b}(-\omega_1,\boldsymbol{k}-\boldsymbol{k}_1)\rangle\langle u_{w,c}(\omega_2,\boldsymbol{k}_2)u_{w,d}(\omega-\omega_2,-\boldsymbol{k}-\boldsymbol{k}_2)\rangle\ \\
&+ \langle(u_{w,a}(\omega_1,\boldsymbol{k}_1)  u_{w,c}(\omega_2,\boldsymbol{k}_2)\rangle \langle u_{w,b}(-\omega_1,\boldsymbol{k}-\boldsymbol{k}_1)u_{w,d}(\omega-\omega_2,-\boldsymbol{k}-\boldsymbol{k}_2)\rangle\\
&+ \langle(u_{w,a}(\omega_1,\boldsymbol{k}_1) u_{w,d}(\omega-\omega_2,-\boldsymbol{k}-\boldsymbol{k}_2) \rangle \langle u_{w,b}(-\omega_1,\boldsymbol{k}-\boldsymbol{k}_1)u_{w,c}(\omega_2,\boldsymbol{k}_2)\rangle.
\end{align}
As a stationary process, each two-point correlation function vanishes unless it has opposing frequencies (e.g. $\omega=-\omega'$).
Likewise, spatial translation invariance means that two-point function vanishes unless it has opposing wave-vectors.
Examining the three terms, we see that all but the last requires $\boldsymbol{k}=0$.
These do not contribute because, as we shall see, $\mathcal{L}(\omega,0) = 0$\footnote{Physically this arises because, in a stratified medium, there must be diffusion to permit motion and that does not happen for the $\boldsymbol{k}=0$ mode.}.
As a result we find
\begin{align}
	D &= \frac{1}{2} \int_{-\infty}^{\infty} d\omega \int_{-\infty}^{\infty} d\omega_1\int_{-\infty}^{\infty} d\omega_2 \sum_{\boldsymbol{k},\boldsymbol{k}_1,\boldsymbol{k}_2} \mathcal{L}_{a}(0,\boldsymbol{k})  \mathcal{L}_{c}(\omega,-\boldsymbol{k}) (\boldsymbol{k}-\boldsymbol{k}_1)_a (-\boldsymbol{k}-\boldsymbol{k}_2)_c \\
	&\times\langle(u_{w,a}(\omega_1,\boldsymbol{k}_1) u_{w,d}(\omega-\omega_2,-\boldsymbol{k}-\boldsymbol{k}_2) \rangle \langle u_{w,b}(-\omega_1,\boldsymbol{k}-\boldsymbol{k}_1)u_{w,c}(\omega_2,\boldsymbol{k}_2)\rangle.
\end{align}
The correlation functions all vanish unless their wave-vectors sum to zero, so $\boldsymbol{k}-\boldsymbol{k}_1+\boldsymbol{k}_2=0$, which we can use to eliminate $\boldsymbol{k}_2$ and find
\begin{align}
	D &= \frac{1}{2} \int_{-\infty}^{\infty} d\omega \int_{-\infty}^{\infty} d\omega_1\int_{-\infty}^{\infty} d\omega_2 \sum_{\boldsymbol{k},\boldsymbol{k}_1} \mathcal{L}_{a}(0,\boldsymbol{k})  \mathcal{L}_{c}(\omega,-\boldsymbol{k}) (\boldsymbol{k}-\boldsymbol{k}_1)_a (-\boldsymbol{k}_1)_c \\
	&\times\langle(u_{w,a}(\omega_1,\boldsymbol{k}_1) u_{w,d}(\omega-\omega_2,-\boldsymbol{k}_1) \rangle \langle u_{w,b}(-\omega_1,\boldsymbol{k}-\boldsymbol{k}_1)u_{w,c}(\omega_2,\boldsymbol{k}_1-\boldsymbol{k})\rangle.
\end{align}
We can shift $\omega$ up by $\omega_2$ and $\boldsymbol{k}$ up by $\boldsymbol{k}_1$ to obtain
\begin{align}
	D &= -\frac{1}{2} \int_{-\infty}^{\infty} d\omega \int_{-\infty}^{\infty} d\omega_1\int_{-\infty}^{\infty} d\omega_2 \sum_{\boldsymbol{k},\boldsymbol{k}_1} \mathcal{L}_{a}(0,\boldsymbol{k}+\boldsymbol{k}_1)  \mathcal{L}_{c}(\omega+\omega_2,-\boldsymbol{k}-\boldsymbol{k}_1) k_a k_{1,c} \\
	&\times\langle(u_{w,a}(\omega_1,\boldsymbol{k}_1) u_{w,d}(\omega,-\boldsymbol{k}_1) \rangle \langle u_{w,b}(-\omega_1,\boldsymbol{k})u_{w,c}(\omega_2,-\boldsymbol{k})\rangle
\end{align}
Inserting equation~\eqref{eq:S2} twice we find
\begin{align}
	D &= -\frac{1}{2} \int_{-\infty}^{\infty} d\omega \int_{-\infty}^{\infty} d\omega_1\int_{-\infty}^{\infty} d\omega_2 \sum_{\boldsymbol{k},\boldsymbol{k}_1} \mathcal{L}_{a}(0,\boldsymbol{k}+\boldsymbol{k}_1)  \mathcal{L}_{c}(\omega+\omega_2,-\boldsymbol{k}-\boldsymbol{k}_1) k_a k_{1,c} \\
	&\times \delta(\omega_1+\omega) S_{w,ad}(\omega_1,\boldsymbol{k}_1) \delta(\omega_2-\omega_1) S_{w,cb}(-\omega_1,\boldsymbol{k}),
\end{align}
where $S_{w}$ is the power spectrum of the wave velocity.
Evaluating the integrals yields
\begin{align}
	D &= -\frac{1}{2} \int_{-\infty}^{\infty} d\omega_1 \sum_{\boldsymbol{k},\boldsymbol{k}_1} \mathcal{L}_{a}(0,\boldsymbol{k}+\boldsymbol{k}_1)  \mathcal{L}_{c}(0,-\boldsymbol{k}-\boldsymbol{k}_1) k_a k_{1,c}  S_{w,ad}(\omega_1,\boldsymbol{k}_1) S_{w,cb}(-\omega_1,\boldsymbol{k})\\
	&= -\frac{1}{2} \int_{-\infty}^{\infty} d\omega_1 \sum_{\boldsymbol{k},\boldsymbol{k}_1} \left(\boldsymbol{k}\cdot\overleftrightarrow{S}_{w}(\omega_1,\boldsymbol{k}_1)\cdot\mathcal{L}(0,\boldsymbol{k}+\boldsymbol{k}_1)\right)\left(\boldsymbol{k}_1\cdot \overleftrightarrow{S}_{w}(-\omega_1,\boldsymbol{k})\cdot  \mathcal{L}(0,-\boldsymbol{k}-\boldsymbol{k}_1)\right).
\end{align}
That is, the diffusivity is given by a bilinear function of the wave power spectrum.

Negating $\boldsymbol{k}_1$ we find
\begin{align}
	D &= \frac{1}{2} \int_{-\infty}^{\infty} d\omega_1 \sum_{\boldsymbol{k},\boldsymbol{k}_1} \left(\boldsymbol{k}\cdot\overleftrightarrow{S}_{w}(\omega_1,-\boldsymbol{k}_1)\cdot\mathcal{L}(0,\boldsymbol{k}-\boldsymbol{k}_1)\right)\left(\boldsymbol{k}_1\cdot \overleftrightarrow{S}_{w}(-\omega_1,\boldsymbol{k})\cdot  \mathcal{L}(0,\boldsymbol{k}_1-\boldsymbol{k})\right).
\end{align}
We can clean this up a little by noting that for real-valued velocity fields $S(\omega,\boldsymbol{k})=S(\pm \omega,\pm \boldsymbol{k})$. So
\begin{align}
	D &= \frac{1}{2} \int_{-\infty}^{\infty} d\omega_1 \sum_{\boldsymbol{k},\boldsymbol{k}_1} \left(\boldsymbol{k}\cdot\overleftrightarrow{S}_{w}(\omega_1,\boldsymbol{k}_1)\cdot\mathcal{L}(0,\boldsymbol{k}-\boldsymbol{k}_1)\right)\left(\boldsymbol{k}_1\cdot \overleftrightarrow{S}_{w}(\omega_1,\boldsymbol{k})\cdot  \mathcal{L}(0,\boldsymbol{k}_1-\boldsymbol{k})\right)\\
	&= \frac{1}{2} \int_{-\infty}^{\infty} d\omega \sum_{\boldsymbol{k}_1,\boldsymbol{k}_2} \left(\boldsymbol{k}_1\cdot\overleftrightarrow{S}_{w}(\omega_1,\boldsymbol{k}_2)\cdot\mathcal{L}(0,\boldsymbol{k}_1-\boldsymbol{k}_2)\right)\times\left(\boldsymbol{k}_1\leftrightarrow \boldsymbol{k}_2\right).
	\label{eq:Dsketch}
\end{align}
where in the last line we have also relabeled $\omega_1 \rightarrow \omega$, $\boldsymbol{k} \rightarrow k_1$, and $\boldsymbol{k}_1 \rightarrow \boldsymbol{k}_2$.

\subsection{Filling in Details}

We now fill in the details we omitted above.
Given the non-linear forcing $\boldsymbol{a}_s$, how does the velocity field respond?
In Appendix~\ref{appen:eom} we derive the linearized equations of motion for IGW in the Boussinesq plane-parallel limit.
We denote Eulerian perturbations by a prime, so that the perturbation of quantity $A$ is written as $A'$, and we write the unperturbed background quantities with a subscript $0$, as in $A_0$.
The subscript $r$ denotes the vertical direction, and $h$ denotes the horizontal one.
Gravity is in the vertical direction.
With this, we obtain equations~\eqref{eq:FT0}-\eqref{eq:FT4}:
\begin{align}
i\omega\rho_0 u_r -i k_r p' - T'\frac{g_0\rho_0 }{T_0} + \mu' \frac{g_0\rho_0 }{\mu_0} &= 0\\
i\omega\rho_0 \boldsymbol{u}_h - i \boldsymbol{k}_h p' &= 0\\
k_r u_r + \boldsymbol{k}_h\cdot\boldsymbol{u}_h &= 0\\
(i\omega +\alpha k^2)T'+ \frac{N_T^2  T_0}{ g_0} u_r&=0\\
(i\omega +D_\mu k^2)\mu'  - \frac{N_\mu^2 \mu_0}{g_0} u_r&=0
\end{align}
Here $\mu$ is the mean molecular weight, $T$ is the temperature, $\rho$ is the density, $g > 0$ is the downward acceleration due to gravity, $N_T$ is the thermal buoyancy frequency, $\alpha$ is the thermal diffusivity, $N_\mu$ is the compositional buoyancy frequency, and $D_\mu$ is the compositional diffusivity.
Note that we work in Fourier space, with wave-vector $\boldsymbol{k}$ and frequency $\omega$.

We can insert our forcing term $\boldsymbol{a}_s$ on the right-hand side of the momentum equation, giving
\begin{align}
i\omega\rho_0 u_r -i k_r p' - T'\frac{g_0}{T_0} + \mu' \frac{g_0}{\mu_0} &= -a_{s,r}\\
i\omega\rho_0 \boldsymbol{u}_h - i \boldsymbol{k}_h p' &= -\boldsymbol{a}_{s,h}.
\end{align}
Solving for the vertical velocity $u_r$ at $\omega=0$ we obtain
\begin{align}
u_r = \frac{\alpha D_\mu k^2}{D_\mu N_T^2 + \alpha N_\mu^2}\left(a_{s,r} - \boldsymbol{a}_{s,\perp}\cdot\boldsymbol{k}_\perp \frac{k_r}{k_\perp^2}\right)
\label{eq:ur}
\end{align}
With this, we find
\begin{align}
	\mathcal{L}(0,\boldsymbol{k}) = \frac{\alpha D_\mu k^2}{D_\mu N_T^2 + \alpha N_\mu^2}\left(\hat{r} - \boldsymbol{k}_\perp \frac{k_r}{k_\perp^2}\right).
\end{align}
Inserting this into equation~\eqref{eq:Dsketch} we find
\begin{align}
	D &= \frac{\alpha^2 D_\mu^2}{2(D_\mu N_T^2 + \alpha N_\mu^2)^2} \int_{-\infty}^{\infty} d\omega \sum_{\boldsymbol{k}_1,\boldsymbol{k}_2} |\boldsymbol{k}_1-\boldsymbol{k}_2|^4 \left(\boldsymbol{k}_1\cdot\overleftrightarrow{S}_{w}(\omega_1,\boldsymbol{k}_2)\cdot\left[\hat{r} - (\boldsymbol{k}_{1,\perp}-\boldsymbol{k}_{2,\perp})\frac{k_{1,r}-k_{2,r}}{|\boldsymbol{k}_{1,\perp}-\boldsymbol{k}_{2,\perp}|^2}\right]\right)\times\left(\boldsymbol{k}_1\leftrightarrow \boldsymbol{k}_2\right).
	\label{eq:Dfinal}
\end{align}

\subsection{Approximate Expression}

Equation~\eqref{eq:Dfinal} is rather unwieldy.
It may be simplified by noting that $k_r \gg k_\perp$ and $k_r u_{w,r} \approx k_\perp u_{w,\perp}$ for IGW.
This means that the $\hat{r}$ term in $\mathcal{L}$ contributes very little and that $k \approx k_r$, so
\begin{align}
	D &\approx \frac{\alpha^2 D_\mu^2}{2(D_\mu N_T^2 + \alpha N_\mu^2)^2} \int_{-\infty}^{\infty} d\omega \sum_{\boldsymbol{k}_1,\boldsymbol{k}_2} \frac{|\boldsymbol{k}_1-\boldsymbol{k}_2|^6}{|\boldsymbol{k}_{1,\perp}-\boldsymbol{k}_{2,\perp}|^4} \left(\boldsymbol{k}_1\cdot\overleftrightarrow{S}_{w}(\omega_1,\boldsymbol{k}_2)\cdot\left[\boldsymbol{k}_{1,\perp}-\boldsymbol{k}_{2,\perp}\right]\right)\times\left(\boldsymbol{k}_1\leftrightarrow \boldsymbol{k}_2\right).
\end{align}
If the spectrum peaks strongly at frequency $\omega \approx \omega_0$ with width $\Delta \omega \approx \omega_0$, and peaks at wave-vector $k_{\perp} \approx k_{\perp,0}$ with width $\Delta k_\perp \approx k_{\perp,0}$, then
 \begin{align}
	D &\approx \frac{\alpha^2 D_\mu^2 k_{r,0}^2}{2 \omega_0 (D_\mu N_T^2 + \alpha N_\mu^2)^2} \left(\frac{k_{r,0}}{k_{\perp,0}}\right)^4 \left(k_{r,0} k_{\perp,0} u_{w,r} u_{w,\perp}\right)^2.
\end{align}
Using the incompressibility condition we find $u_r \approx u_\perp k_\perp / k_r$ so
 \begin{align}
	D &\approx \frac{\alpha^2 D_\mu^2 k_{r,0}^2}{2 \omega_0 (D_\mu N_T^2 + \alpha N_\mu^2)^2} \left(\frac{k_{r,0}}{k_{\perp,0}}\right)^4 \left(k_{\perp,0}^2 u_{w,\perp}^2\right)^2.
\end{align}
It is often convenient to write this in terms of the wave luminosity $L_{w} \approx 4\pi r^2 \rho (\omega/k_r) u_\perp^2$, so
  \begin{align}
	D &\approx \frac{\alpha^2 D_\mu^2 k_{r,0}^8}{2 \omega_0^3 (D_\mu N_T^2 + \alpha N_\mu^2)^2} \left(\frac{L_{\rm wave} }{4\pi r^2 \rho}\right)^2.
	\label{eq:approx}
\end{align}

\section{Discussion}\label{sec:conclusions}

We have derived the leading order non-linear wave mixing due to internal gravity waves in a thermally and compositionally-stratified fluid.
We find that this occurs at fourth order in the wave velocity, scales strongly with both the thermal and compositional diffusivities, and is suppressed by both forms of stratification.

A different expression was obtained by~\citet{1991ApJ...377..268G} by assuming that waves drive shear turbulence which then produces mixing. That expression is linear in the wave luminosity (quadratic in the velocity), linear in the thermal diffusivity, and generally predicts much more mixing than our expression.

We have not studied the wave-driven turbulence scenario, but note that in order for this to produce substantial mixing it must mean that a large fraction of the wave power is processed into zero-frequency motion.
We encourage further study of whether and how this happens to pin down the scaling of wave mixing.

\acknowledgments

I am grateful to Jim Fuller and Yuri Levin for extensive discussions, feedback, and mentorship on this work.
The Flatiron Institute is supported by the Simons Foundation.
This work was also supported by the Gordon and Betty Moore Foundation (Grant GBMF7392) and the National Science Foundation (Grant No. NSF PHY-1748958).

\clearpage

\appendix

\section{Equations of Motion}\label{appen:eom}

Here we derive the linearized equations of motion for IGW in the Boussinesq plane-parallel limit, taking inspiration from~\citet{ChristensenDalsgaard2003}.

\subsection{Mass Equation}

In the Boussinesq approximation we neglect density perturbations except in the momentum equation, so the continuity equation for mass is
\begin{align}
\boldsymbol{u}\cdot\nabla\rho_0 + \rho_0 \nabla\cdot\boldsymbol{u} = 0.
\end{align}
In this approximation we further neglect the background density gradient, assuming the waves to have a much smaller characteristic vertical scale, so this reduces to
\begin{align}
\nabla\cdot\boldsymbol{u} = 0.
\label{eq:mass}
\end{align}

\subsection{Composition Equation}

We treat composition via the mean molecular weight $\mu$, which follows an advection-diffusion equation
\begin{align}
	\partial_t \mu + u_r \partial_r \mu - D_\mu \nabla^2 \mu = 0,
\end{align}
where $D_\mu$ is the compositional diffusivity and $\partial_r$ is the vertical spatial derivative.
Note that we have already made use of the Boussinesq approximation by neglecting density variation, and we have assumed that $D_\mu$ is a constant so that it commutes with $\nabla$.

Expanding this equation to linear order in the perturbations we find
\begin{align}
\partial_t \mu' + u_r' \partial_r \mu_0 - D_\mu \nabla^2 \mu' = 0
\end{align}
where we have assumed $u_{r,0}=0$, corresponding to a stationary background state.
Defining
\begin{align}
N_\mu^2 \equiv -\frac{g_0}{\mu_0} \partial_r \mu_0,
\end{align}
where $g_0$ is the background acceleration due to gravity, we find
\begin{align}
\partial_t \mu' - u_r \frac{N_\mu^2 \mu_0}{g_0} - D_\mu \nabla^2 \mu' = 0
\label{eq:comp}
\end{align}

\subsection{Energy Equation}

The energy equation is
\begin{align}
	c_p \partial_t T + u_r c_p T \partial_r s = -\nabla\cdot\boldsymbol{F},
\end{align}
where $s$ is the dimensionless entropy, $\boldsymbol{F}$ is the radiative heat flux, and $c_p$ is the specific heat at constant pressure.
Here we have assumed that the entropy is constant in the horizontal direction.

We can expand the entropy gradient in terms of the temperature gradient as
\begin{align}
	T \partial_r s = \partial_r T - \partial_r T_{\rm ad},
\end{align}
where the second term on the right-hand side is the adiabatic temperature gradient.
This gives
\begin{align}
	c_p \partial_t T + u_r c_p \partial_r (T - T_{\rm ad}) = -\nabla\cdot\boldsymbol{F},
\end{align}
Next, we write the heat flux as
\begin{align}
	\boldsymbol{F} = -\alpha c_p \nabla T,
\end{align}
where $\alpha$ is the thermal diffusivity.
Treating $\alpha$ and $c_p$ as constants we find
\begin{align}
	c_p \partial_t T + u_r c_p \partial_r (T - T_{\rm ad}) = \alpha c_p \nabla^2 T.
\end{align}
Expanding to linear order, we see that
\begin{align}
	c_p \partial_t T' + u_r c_p \partial_t (T_0 - T_{\rm ad}) = \alpha c_p \nabla^2 T'
\end{align}
Defining
\begin{align}
	N_T^2 \equiv \frac{g_0}{T_0}\partial_r (T_0 - T_{\rm ad}),
\end{align}
we finally write this as
\begin{align}
	\partial_t T' + \frac{u_r}{g_0} T_0 N_T^2 = \alpha \nabla^2 T'.
	\label{eq:energy}
\end{align}

\subsection{Momentum Equation}

The inviscid linearized Boussinesq Navier Stokes Equation is
\begin{align}
\rho_0 \partial_t \boldsymbol{u} = -\nabla p' + \rho_0 \boldsymbol{g}' + \rho' \boldsymbol{g}_0.
\end{align}
Neglecting the perturbation to the gravitational field we find
\begin{align}
\rho_0 \partial_t \boldsymbol{u} = -\nabla p' + \rho' \boldsymbol{g}_0+\rho_0\nu\nabla^2 \boldsymbol{u}.
\end{align}
Expanding the density perturbation in terms of the composition and temperature we obtain
\begin{align}
\rho_0 \partial_t \boldsymbol{u} = -\nabla p' + \rho_0 \left(\frac{\mu'}{\mu_0}-\frac{T'}{T_0}\right) \boldsymbol{g}_0+\rho_0\nu\nabla^2 \boldsymbol{u}.
\end{align}
Splitting this into a horizontal component and a radial component we find
\begin{align}
\rho_0 \partial_t \boldsymbol{u}_h - \rho_0 \nu \nabla^2 \boldsymbol{u}_h&= -\nabla_h p'\\
\rho_0 \partial_t u_r- \rho_0 \nu \nabla^2 u_r &= -\partial_r p' - g_0 \rho_0 \left(\frac{\mu'}{\mu_0}-\frac{T'}{T_0}\right) 
\end{align}
where we have picked a sign convention such that $\boldsymbol{g}_0$ points radially downward and the scalar $g_0 > 0$.

\subsection{Fourier Transform}
Suppose that our solution is proportional to $e^{i\omega t - i k_h x_h-ik_r r}$. Then our equations become
\begin{align}
\label{eq:FT0}
i\omega\rho_0 u_r -i k_r p' - T'\frac{g_0}{T_0} + \mu' \frac{g_0}{\mu_0} &= 0\\
\label{eq:FT1}
i\omega\rho_0 \boldsymbol{u}_h - i \boldsymbol{k}_h p' &= 0\\
\label{eq:FT2}
k_r u_r + \boldsymbol{k}_h\cdot\boldsymbol{u}_h &= 0\\
\label{eq:FT3}
(i\omega +\alpha k^2)T'+ \frac{N_T^2  T_0}{ g_0} u_r&=0\\
\label{eq:FT4}
(i\omega +D_\mu k^2)\mu'  - \frac{N_\mu^2 \mu_0}{g_0} u_r&=0
\end{align}

\bibliography{refs}
\bibliographystyle{aasjournal}

\end{document}